\documentclass[sigconf,nonacm]{acmart}

\usepackage{amsmath}
\IfFileExists{newtxmath.sty}{}{\usepackage{amssymb,amsfonts}}
\usepackage{array}
\usepackage{placeins}
\usepackage{stfloats}
\usepackage[most]{tcolorbox}
\microtypesetup{expansion=false}
\hypersetup{hidelinks}

\makeatletter
\newsavebox{\firstpagecomparisonbox}
\newcommand{\placefirstpagecomparison}{%
  \AddToHookNext{cmd/@topnewpage/after}{%
    \global\advance\@colht-\ht\firstpagecomparisonbox
    \global\advance\@colht-\dp\firstpagecomparisonbox
    \global\advance\@colht-8pt
    \global\@colroom\@colht
    \global\vsize\@colht}%
  \AddToHookNext{cmd/@combinedblfloats/after}{%
    \ifnum\value{page}=1\else
      \PackageError{pageone-layout}{Comparison table must be on page 1}{Place the comparison box before maketitle.}%
    \fi
    \setbox\@outputbox=\vbox to\textheight{%
      \unvbox\@outputbox
      \vskip8pt
      \nointerlineskip
      \copy\firstpagecomparisonbox
      \vskip0pt}}}
\AddToHook{begindocument/end}{%
  \let\pageone@section\section
  \let\pageone@textbottom\@textbottom
  \let\pageone@texttop\@texttop
  \edef\pageone@textfloatsep{\the\textfloatsep}%
  \raggedbottom
  \setlength{\textfloatsep}{6pt}%
  \patchcmd{\section}
    {-.75\baselineskip \@plus -2\p@ \@minus -.2\p@}
    {-3pt}{}%
    {\PackageError{pageone-layout}{Section spacing patch failed}{Check the acmart section definition.}}%
  \AddToHookNext{shipout/after}{%
    \global\let\section\pageone@section
    \global\let\@textbottom\pageone@textbottom
    \global\let\@texttop\pageone@texttop
    \global\textfloatsep=\pageone@textfloatsep\relax}}
\makeatother

\title[HBF Sucks?]{HBF Sucks? A Full-Stack Characterization of High-Bandwidth Flash for KV-Centric LLM Serving}

\author{Zhuoran Li}
\affiliation{
  \department{School of Integrated Circuits}
  \institution{Peking University}
  \city{Beijing}
  \country{China}}
\email{zhuoranli@stu.pku.edu.cn}

\author{Zhuohang Bian}
\affiliation{
  \department{School of Integrated Circuits}
  \institution{Peking University}
  \city{Beijing}
  \country{China}}
\email{bianzhuohang26@stu.pku.edu.cn}

\author{Xin Huang}
\affiliation{
  \department{Fudan Institute of Systems for Advanced Computing}
  \institution{Fudan University}
  \city{Shanghai}
  \country{China}}
\email{huangxin25@m.fudan.edu.cn}

\author{Yibo Zhao}
\affiliation{
  \department{School of Integrated Circuits}
  \institution{Peking University}
  \city{Beijing}
  \country{China}}
\email{yibozhao@stu.pku.edu.cn}

\author{Guangyu Sun}
\affiliation{
  \department{School of Integrated Circuits}
  \institution{Peking University}
  \city{Beijing}
  \country{China}}
\email{gsun@pku.edu.cn}

\author{Youwei Zhuo}
\affiliation{
  \department{School of Integrated Circuits}
  \institution{Peking University}
  \city{Beijing}
  \country{China}}
\email{youwei@pku.edu.cn}

\makeatletter
\renewcommand{\@mkauthors@iii}{%
  \gdef\@currentauthors{}%
  \global\setbox\mktitle@bx=\vbox{%
    \unvbox\mktitle@bx\par\smallskip
    \centering
    {\large\sffamily
      Zhuoran Li\textsuperscript{1}\quad
      Zhuohang Bian\textsuperscript{1}\quad
      Xin Huang\textsuperscript{2}\quad
      Yibo Zhao\textsuperscript{1}\quad
      Guangyu Sun\textsuperscript{1}\quad
      Youwei Zhuo\textsuperscript{1,*}\par}
    \smallskip
    {\normalsize
      \textsuperscript{1}School of Integrated Circuits, Peking University, Beijing, China\par
      \textsuperscript{2}Fudan Institute of Systems for Advanced Computing, Fudan University, Shanghai, China\par}
    {\small
      \{\href{mailto:zhuoranli@stu.pku.edu.cn}{zhuoranli},
      \href{mailto:bianzhuohang26@stu.pku.edu.cn}{bianzhuohang26},
      \href{mailto:yibozhao@stu.pku.edu.cn}{yibozhao}\}@stu.pku.edu.cn\quad
      \{\href{mailto:gsun@pku.edu.cn}{gsun},
      \href{mailto:youwei@pku.edu.cn}{youwei}\}@pku.edu.cn\quad
      \href{mailto:huangxin25@m.fudan.edu.cn}{huangxin25@m.fudan.edu.cn}\quad
      \textsuperscript{*}Corresponding author\quad
      \textit{Simulator}:\ \url{https://github.com/pku-lemonade/TokenSim/tree/hbf}\par}
    \medskip}}
\makeatother

\renewcommand{\shortauthors}{Li et al.}

\definecolor{takeawayblue}{RGB}{47,84,120}
\definecolor{takeawayfill}{RGB}{235,242,249}
\newtcolorbox{findingtakeaway}[1]{
  enhanced,
  colback=takeawayfill,
  colframe=takeawayblue,
  colbacktitle=takeawayblue,
  coltitle=white,
  title={#1},
  fonttitle=\normalsize\bfseries,
  fontupper=\normalsize\bfseries,
  boxrule=0.6pt,
  arc=0.8mm,
  left=1.6mm,
  right=1.6mm,
  top=0.7mm,
  bottom=0.7mm,
  toptitle=0.4mm,
  bottomtitle=0.4mm,
  before skip=5pt,
  after skip=5pt
}

\newtcolorbox{readerchallenge}[1]{
  enhanced,
  colback=takeawayfill,
  colframe=takeawayblue,
  colbacktitle=takeawayblue,
  coltitle=white,
  title={#1},
  fonttitle=\normalsize\bfseries,
  fontupper=\small,
  boxrule=0.6pt,
  arc=0.8mm,
  left=1.8mm,
  right=1.8mm,
  top=1.0mm,
  bottom=1.0mm,
  toptitle=0.5mm,
  bottomtitle=0.5mm,
  before skip=0pt,
  after skip=0pt
}

\begin{document}
\begin{teaserfigure}
  \begin{readerchallenge}{A Challenge to the Reader}
    Every clue is fairly presented below: the traces, models, code, and three necessary conditions.
    Mooncake-style KV offload fails all three---but HBF need not. This paper already supplies part of the
    solution. Can you complete the design---an HBF organization and runtime that accelerates critical-path
    reads, earns enough useful reads per write, sustains bandwidth within thermal and endurance limits, and
    delivers an end-to-end win?
  \end{readerchallenge}
  \Description{A challenge to the reader to complete an HBF organization and runtime design.}
\end{teaserfigure}

\begin{abstract}
  A faster storage device should make serving faster. We find the opposite. High-Bandwidth Flash (HBF) stacks
  NAND behind a wide, package-local interface, promising flash-scale capacity with far lower read latency and
  higher bandwidth than an SSD. The obvious move is to keep an SSD-style Mooncake KV-offloading stack and swap in
  HBF underneath. We built that system and measured it: an extended TokenSim, four complete two-hour
  Qwen-Bailian production traces, five dense and mixture-of-experts models, and H100/B200 profiles. The upgrade
  backfires. Average end-to-end latency rises 2--5.5$\times$ and maximum SLO goodput falls 1.1--2.7$\times$
  across H100 and B200, so the faster device yields a slower system. A cost-benefit model explains the paradox: a
  faster far tier pays off only when read I/O is the bottleneck, reads outweigh writes, and delivered bandwidth
  is sustainable. Transient KV violates all three at once. Buying flash through the package costs GPU near-tier
  capacity and bandwidth, while HBF's own read/write latency barely matters: scaling it 3.75$\times$ moves
  latency less than 1\%. Worse, the two-tier hierarchy keeps reuse in the near tier and hands HBF a relentless
  write-heavy stream. Writes outnumber reads on every trace, so a 3D-ICE model shows the stack hits its thermal
  limit well below peak bandwidth, and a TLC tier wears out sooner than the SSD pool it replaced. The device is
  fine; the drop-in deployment is not. HBF sucks as an SSD replacement for transient KV, but earns its place in
  LLM serving when used selectively with reuse-aware placement, write budgeting, and thermal coordination.
  \end{abstract}

\begin{lrbox}{\firstpagecomparisonbox}
\begin{minipage}[b]{\textwidth}
  \small
  \setlength{\tabcolsep}{1.5pt}
  \setlength{\aboverulesep}{0.25ex}
  \setlength{\belowrulesep}{0.35ex}
  \renewcommand{\arraystretch}{0.9}
  \captionsetup{type=table,skip=3pt}
  \caption{HBF topology and evaluation comparison~\cite{Kim2026HBFRoadmap}.}
  \label{tab:hbf-work-comparison}
  \noindent\begin{tabular}{@{}
    >{\raggedright\arraybackslash}p{\dimexpr0.14\textwidth-1.6\tabcolsep\relax}
    >{\raggedright\arraybackslash}p{\dimexpr0.23\textwidth-1.6\tabcolsep\relax}
    >{\raggedright\arraybackslash}p{\dimexpr0.16\textwidth-1.6\tabcolsep\relax}
    >{\raggedright\arraybackslash}p{\dimexpr0.22\textwidth-1.6\tabcolsep\relax}
    >{\raggedright\arraybackslash}p{\dimexpr0.25\textwidth-1.6\tabcolsep\relax}@{}}
    \toprule
    \textbf{Work} & \textbf{HBF-resident data} & \textbf{HBF topology} &
    \textbf{Architecture simulator} & \textbf{Serving evaluation} \\
    \midrule
    H\textsuperscript{3}~\cite{HBFHa2026H3} & Read-only weights + shared KV & HBF-3 &
    Analytical & Controlled batches \\
    \midrule[0.2pt]
    FlashAccel~\cite{Wang2026FlashAccel} & Weights + KV cache & HBF-2 / HBF-3 &
    Hybrid model & Controlled sweeps \\
    \midrule[0.2pt]
    FLINT~\cite{Oliveira2026FLINT} & Read-only model weights & HBF-3 &
    Trace-coupled & Trace-derived decode \\
    \midrule[0.2pt]
    DASH~\cite{Kim2026DASH} & Expert weights + dynamic KV & HBF-2 + HBF-3 &
    Analytical & Controlled batching \\
    \midrule[0.2pt]
    HBFSim~\cite{Hu2026HBFSim} & GPU data; expert-weight slice & GPU-visible HBF &
    Hardware-coupled & Runtime replay \\
    \midrule[0.2pt]
    Exploring HBF~\cite{Son2026ExploringHBF} & Model weights + KV & HBF-2 / HBF-only &
    Analytical & Controlled batching \\
    \midrule
    \textbf{This work} & \textbf{Transient KV; SSD-style offload} &
    \textbf{HBF-1 + HBF-2} & \textbf{Full-stack TokenSim\newline + 3D-ICE} &
    \textbf{\ding{51} Production replay: 4 $\times$ 2-h Qwen traces} \\
    \bottomrule
  \end{tabular}\par
\end{minipage}
\end{lrbox}
\placefirstpagecomparison
\maketitle

\begin{figure}[t]
  \captionsetup{skip=2pt}
  \centering
  \begin{minipage}[t]{0.49\columnwidth}
    \centering
    \includegraphics[width=\linewidth]{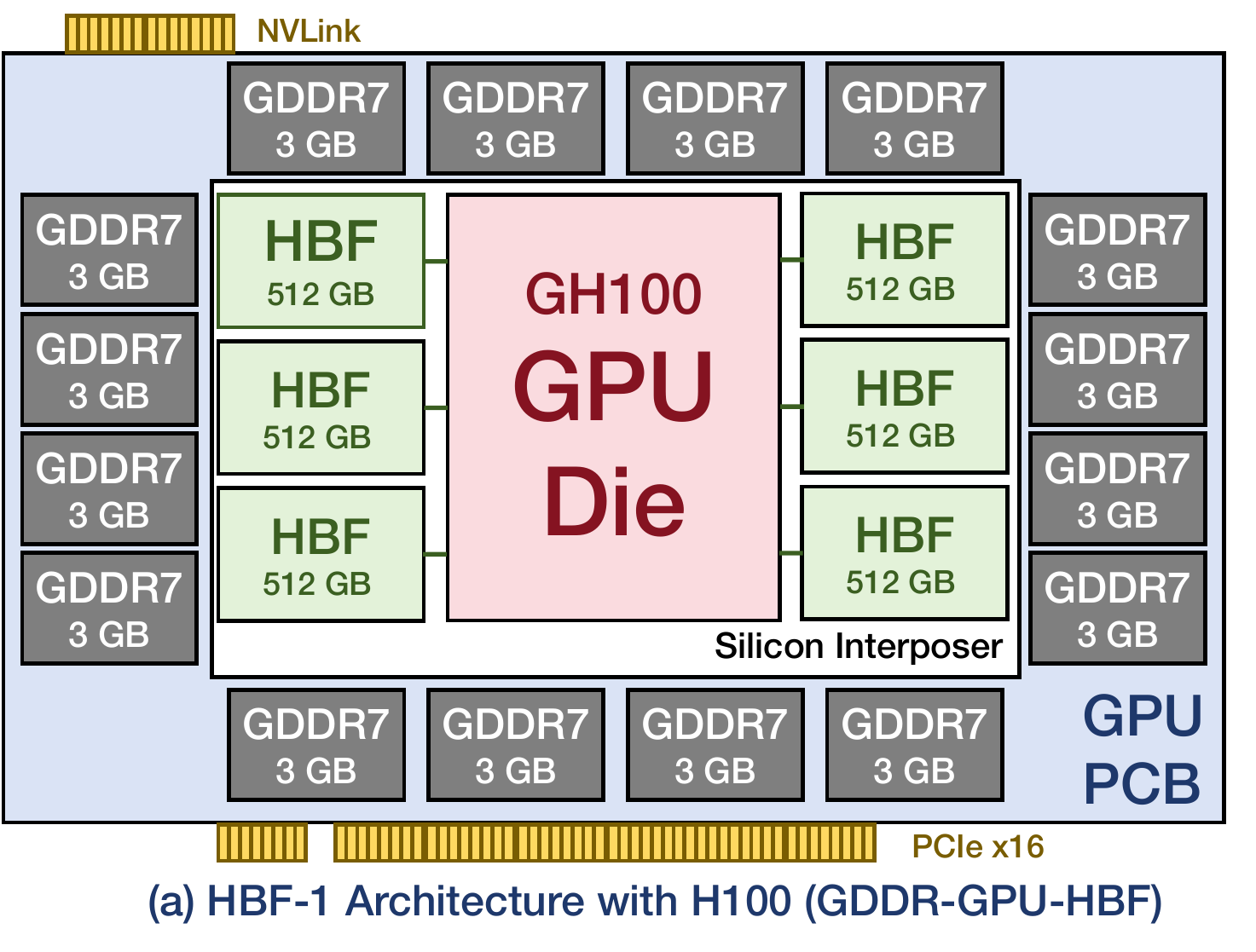}\\[-0.7ex]
    {\footnotesize (a) HBF-1(2028)}
  \end{minipage}\hfill
  \begin{minipage}[t]{0.49\columnwidth}
    \centering
    \includegraphics[width=\linewidth]{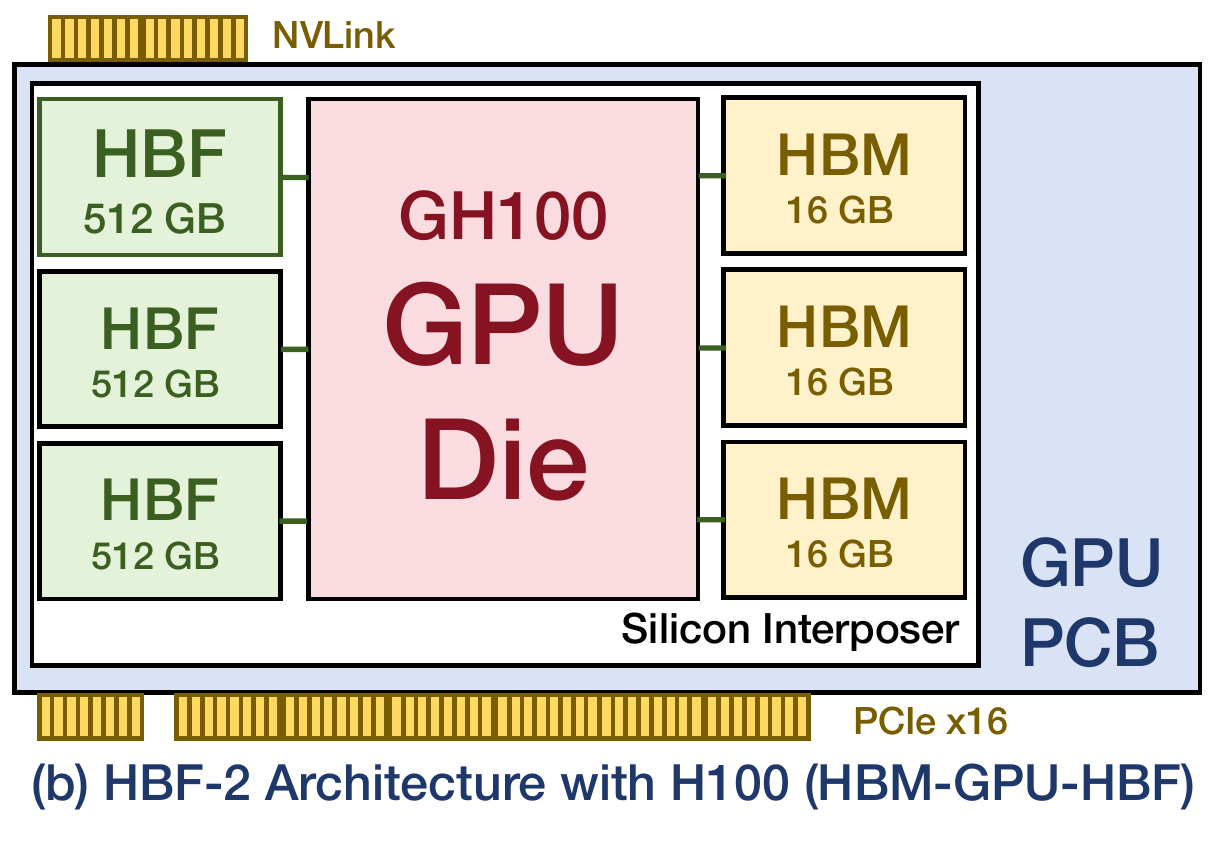}\\[-0.7ex]
    {\footnotesize (b) HBF-2(2030)}
  \end{minipage}
  \caption{Modeled layouts based on the KAIST HBF roadmap.}
  \label{fig:hbf-generations-layout}
  \Description{Side-by-side roadmap system layouts for HBF-1 and HBF-2.}
\end{figure}

\section{Introduction}
\label{sec:introduction}

Large language model (LLM) inference stores a key and value vector for every
retained token. The key--value (KV) cache grows with context length, batch size,
and active requests, limiting concurrency and retained context. Paged allocation,
quantization, and selective retention reduce waste or bytes per
token~\cite{Kwon2023vLLM,Liu2024ICMLKIVI,Hooper2024NeurIPSKVQuant,Zhang2025OSDIDiffKV}
but not the capacity limit. When the working set of prefixes and histories exceeds
GPU memory, the system must place KV in a larger tier.

Modern serving systems offload reusable KV to a larger tier. Reuse comes from
repeated dialogue, common prefixes in shared prompts and documents, and context
revisited by agent workflows. Mooncake and related systems therefore extend the KV
pool into CPU DRAM, remote memory, and SSDs~\cite{Kimi2025MoonCake,Bin2024CachedAttention}.
This offloading is standard and essential: without it, a capacity-limited server
must reject requests or recompute evicted KV. It supplies the objects that larger
tiers store and restore.

High-Bandwidth Flash (HBF) connects stacked NAND to the accelerator through a wide
memory interface. We base our evaluation on the HBF roadmap's
near-term organizations, HBF-1 (2028) and HBF-2 (2030)~\cite{Kim2026HBFRoadmap}.
HBF-1 pairs package-local HBF with a GDDR7 near tier; HBF-2 shares package sites
between HBM and HBF. These dates denote roadmap projections; HBF-3 follows in 2032.

HBF studies demonstrate benefits for read-mostly objects and specialized data
paths~\cite{HBFHa2026H3,HBFHsu2026HAVEN}. We take the interface and management
requirements directly from the OCP base-die specification: AXI over UCIe,
page-oriented writes, and host-directed data layout~\cite{OCP2026HBFBaseDie}.
We study HBF under an unmodified Mooncake-style SSD KV-pooling runtime:
admission, retention, lookup, and KV save and restore remain unchanged while
the backing tier and transport change. Faster flash service thus arrives with
changes in near-tier capacity and bandwidth. The pooling policy determines which
objects reach HBF; admission and queueing determine the serving outcome.
Table~\ref{tab:hbf-work-comparison} places this evaluation alongside recent HBF
studies with other organizations and workloads.

We evaluate this roadmap-constrained deployment under the same runtime, comparing
complete HBF-1 and HBF-2 organizations against SSD baselines using
TokenSim, four complete two-hour production traces, five dense and
mixture-of-experts (MoE) models, and H100/B200 profiles. HBF is slower on every
request metric: across H100 and B200, average end-to-end latency rises
$2$--$5.5\times$, and maximum
service-level objective (SLO) goodput falls $1.1$--$2.7\times$. Faster flash barely
helps; changing HBF read/write latency by $3.75\times$ moves average end-to-end
latency by less than $1\%$.
Table~\ref{tab:evaluation} summarizes the evaluation workloads, model profiles,
and system organizations.

The measured behavior arises from the interaction between the SSD-style hierarchy
and the roadmap package tradeoff. The hierarchy sends HBF transient and write-heavy
KV, whose traffic exercises the physical limits of NAND. The boundary is simple: faster
media helps only when secondary-tier service is on the request critical path (C1),
later reads repay each flash write (C2), and delivered bandwidth is sustainable (C3).
\S~\ref{sec:framework} formalizes these as three necessary conditions, each
with a falsifiable requirement tested by two findings.
HBF helps only when all three hold at once, so failing any one sinks the substitution.

Six findings form three matched pairs, each falsifying one condition:

\textbf{C1---Read I/O is the serving bottleneck.} Complete HBF-1 and HBF-2 gain flash capacity by reducing GPU near-tier capacity and bandwidth, so the
near tier admits fewer requests while the flash service HBF accelerates is a small
fraction of the critical path (Finding~1, \S~\ref{sec:capacity}). Base-die attention
compute cannot raise that fraction (Finding~2, \S~\ref{sec:nmp}).

\textbf{C2---Reads outweigh writes.} The two-tier hierarchy serves reuse from the
near tier and sends HBF a write-heavy cold stream, so writes outnumber reads on every
trace (Finding~3, \S~\ref{sec:objects}). The SSD write-batching optimization that might
lower their cost does not transfer (Finding~4, \S~\ref{sec:batching}).

\textbf{C3---Read/write Performance sustains.} That stream drives the stack to its
thermal limit well below peak bandwidth, forcing plane throttling
(Finding~5, \S~\ref{sec:thermal}). Its write volume also exhausts HBF's endurance
budget sooner than the capacity-matched SSD pool's budget
(Finding~6, \S~\ref{sec:cost}).

HBF sucks as a drop-in SSD KV pool under this unmodified runtime. The result
motivates selective, reuse-aware placement; this paper makes four contributions:
\begin{enumerate}
  \item A model of when a faster far-memory tier helps LLM serving, with three
  necessary conditions: read-I/O exposure, reads per write, and sustained
  bandwidth, evaluated on roadmap-derived HBF-1 and HBF-2 organizations.
  It separates media speed from changes in local capacity, bandwidth, and queueing.
  \item Measurements showing why an SSD-style hierarchy sends HBF a write-heavy,
  low-reuse stream, using per-tier byte counters over four production traces.
  \item A 3D-ICE thermal model and a write-endurance budget that turn HBF's power and
  wear pressures into sourced, clearly bounded estimates.
  \item Requirements for a critical-path-aware, reuse-aware, write-budgeted, and
  thermally managed HBF KV tier.
\end{enumerate}

\section{High-Bandwidth Flash}
\label{sec:hbf-background}

We first describe the physical HBF device separately from the software path. The
HBF roadmap supplies system organizations and projected generation
dates~\cite{Kim2026HBFRoadmap}; HAVEN's modeled NAND arrays supply device
latency, energy, bandwidth, and package power~\cite{HBFHsu2026HAVEN}.
The OCP base-die specification v0.7.0, dated August 3, 2026, defines the
host interface and management responsibilities~\cite{OCP2026HBFBaseDie}.

\subsection{HBF Organization and Stacking}

HBF stacks 3D NAND dies behind wide vertical or package-level interconnects. It
also includes a base die or distributed peripheral circuits, many independently
accessible NAND regions, and a package-level link to the GPU. Together, these
components provide much more internal parallelism and interface bandwidth than a
conventional SSD. The key distinction is simple:
\emph{HBF changes how NAND connects to the processor; it does not change NAND into
DRAM.}

The media hierarchy is
\begin{equation}
\small
\text{stack}\rightarrow\text{die}\rightarrow\text{plane}\rightarrow
\text{block}\rightarrow\text{page}\rightarrow\text{cell}.
\end{equation}
Reads sense pages or subarrays. Programs are slower than reads, and erases use
larger blocks. NAND lacks arbitrary in-place overwrites, so invalid pages require
later reclamation.
Aggregate bandwidth comes from concurrency across channels, dies, planes, and
subarrays, not DRAM-like latency for each access.

\begin{figure}[t]
  \centering
  \includegraphics[width=\columnwidth]{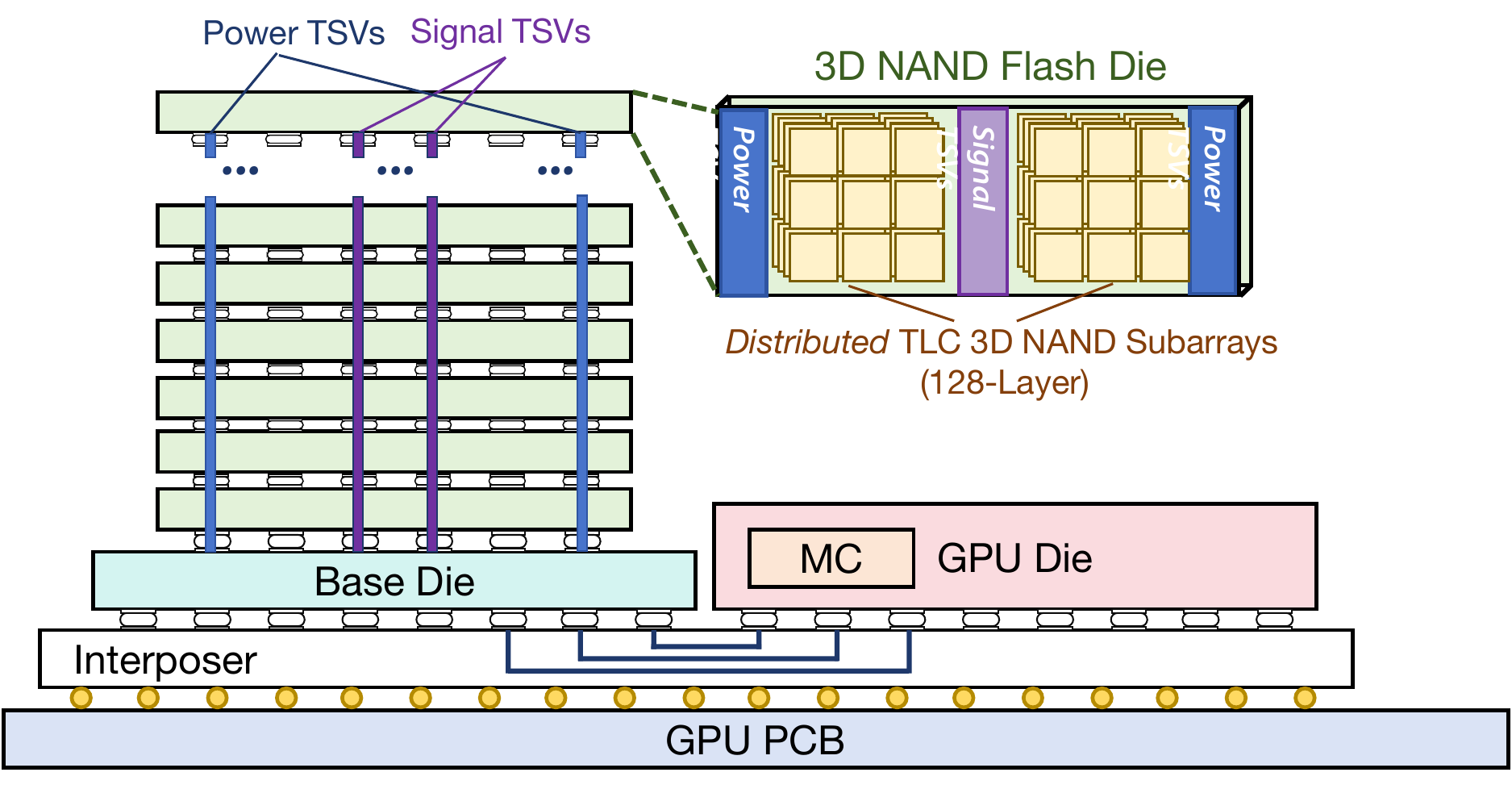}
  \caption{Physical organization of an HBF package.}
  \label{fig:hbf-package}
\end{figure}

A representative stack integrates sixteen 128--256-layer NAND dies,
through-silicon-via (TSV) arrays, and a base die. The base die concentrates mapping
and management logic near the package, a natural control-logic location that also
concentrates area and heat (Fig.~\ref{fig:hbf-package})~\cite{Kim2026HBFRoadmap}.

The OCP specification defines HBF as a non-coherent, memory-centric device using
AXI over UCIe rather than NVMe. The host maps global addresses across independent channels; the base die
maps channel-local addresses to NAND and handles ECC, bad blocks, and command
scheduling~\cite[\S\S3--5]{OCP2026HBFBaseDie}.
OCP specifies 64-byte reads, with two page-cache buffers per bank and ordered
NAND sensing within each bank. Cache hits can bypass sensing, while batch-read
hints control dispatch~\cite[\S5.3]{OCP2026HBFBaseDie}.

OCP requires writes to accumulate a complete 4\,KiB page and complete only after programming the
core die. The host writes pages consecutively within a NAND block; a page-zero
write triggers automatic block erase~\cite[\S5.4]{OCP2026HBFBaseDie}.
OCP also permits an optional host-managed SRAM scratchpad~\cite[\S11.3]{OCP2026HBFBaseDie}.
These mechanisms divide responsibility between the host and base die.
Under the SSD-style pooling path, a later reuse still restores KV to the near
tier before execution, so the pooling policy controls which reads reach NAND.

\subsection{HBF Generations: HBF-1 and HBF-2}

Our HBF-1 and HBF-2 profiles instantiate the HBF roadmap's GDDR--HBF and
HBM--HBF organizations~\cite{Kim2026HBFRoadmap}. Table~\ref{tab:hbf-generations}
lists the modeled H100 instances. Table~\ref{tab:evaluation}(c) also gives the B200
profiles. Stack counts, capacities, and service settings belong to these modeled
profiles, while the organizations and generation dates come from the roadmap.

\begin{table}[t]
  \caption{Modeled H100 instances of KAIST's HBF-1/HBF-2 organizations.}
  \label{tab:hbf-generations}
  \centering
  \small
  \setlength{\tabcolsep}{2.2pt}
  \begin{tabular}{p{0.24\columnwidth}p{0.31\columnwidth}p{0.31\columnwidth}}
    \toprule
    Property & HBF-1 & HBF-2 \\
    \midrule
    Roadmap year & 2028 & 2030 \\
    Organization & Single-GPU GDDR--HBF & Local HBM--HBF \\
    Near tier & 12 GDDR7, 48 GB & 3 HBM, 48 GB \\
    HBF tier & 6 stacks, 3 TB & 3 stacks, 1.5 TB \\
    Weight tier & GDDR7 & HBM \\
    KV movement & On-package GDDR--HBF & Local HBM--HBF \\
    \bottomrule
  \end{tabular}
\end{table}

\subsubsection{HBF-1} (2028) is a single GPU whose silicon interposer is filled entirely with HBF
(Fig.~\ref{fig:hbf-generations-layout}a). The interposer can host only a few stacks, so
once HBF takes those slots no room is left for HBM, and the fast near tier is pushed off
the interposer to GDDR7 in BGA packages on the card PCB. HBF-1 thus buys package-local
flash capacity by demoting its near tier from on-interposer HBM to slower off-interposer
GDDR.

\subsubsection{HBF-2} (2030) shares the interposer between HBM and HBF around one GPU
(Fig.~\ref{fig:hbf-generations-layout}b). Our modeled equal split assigns half the
stack sites to each tier, reducing near-tier capacity and bandwidth relative to
an all-HBM package. HBM holds weights and active, write-heavy state while HBF holds
long-context KV. One scheduler sees compute, HBM, and HBF over the local memory
path, with package area and cooling shared by the two memory tiers.

\subsubsection{HBF-3: a later roadmap organization}
The roadmap places a later HBF-3 generation in 2032, using an advanced interposer
to co-package more HBM and HBF stacks alongside the GPU~\cite{Kim2026HBFRoadmap}.
Our complete-organization evaluation focuses on the earlier HBF-1 and HBF-2
generations.

Because HBF-1 and HBF-2 change GPU organization and near-tier memory with the
medium, a fair study needs two comparisons. The \emph{media-only} comparison varies
only secondary-tier latency, bandwidth, and capacity; the
\emph{complete-architecture} comparison evaluates the full organization. Without
this separation, media service, local memory, interconnect, and package power are
inseparable (Section~\ref{sec:methodology}).

\subsection{Advantages of HBF}

\subsubsection{Capacity and bandwidth}
NAND provides much greater density than DRAM. A 512-GB HBF device paired with a 24-GB HBM stack yields a density ratio of$\approx21.3$.

HBF replaces the narrow PCIe/NVMe path with a wide package interface and exposes
many NAND regions in parallel. The interface is only one bound on effective
application bandwidth:
\begin{multline}
BW_{\mathrm{effective}}=\min(BW_{\mathrm{interface}},BW_{\mathrm{array}},
BW_{\mathrm{controller}},\\
BW_{\mathrm{power}},BW_{\mathrm{thermal}},BW_{\mathrm{workload}}).
\label{eq:effective-bw}
\end{multline}
Peak bandwidth therefore requires enough independent reads, controller concurrency, and power headroom.
The OCP specification makes the layout requirement explicit: the host interleaves
data across channels and aligns bank accesses to expose parallelism~\cite[\S\S11.1--11.2]{OCP2026HBFBaseDie}.

\subsubsection{Shorter path and a cooperative hierarchy}
HBF removes PCIe/NVMe command processing, host staging, and board links. Modern
GDS- and Tutti-style SSD paths narrow this advantage, so we use a strong SSD
baseline. In the intended cooperative hierarchy, HBM holds weights, active KV, and
write-heavy state; HBF holds larger read-mostly reusable objects; and SSD keeps
colder capacity. A NAND read remains slow, but die-, plane-, and channel-level
parallelism provides high aggregate throughput when the runtime exposes concurrency.

\subsection{Limitations of HBF}

The wide interface does not change NAND, so HBF inherits three device-level limits
relative to HBM. First, \emph{latency}: NAND reads and writes take on the order of
microseconds, roughly two orders of magnitude longer than HBM accesses; programs
are slower and more energy-intensive than reads~\cite{HBFHsu2026HAVEN}. Die-,
plane-, and channel-level parallelism raises aggregate bandwidth but does not close
this per-access gap.

Second, \emph{endurance}: flash cells tolerate finite program/erase cycles, and
elevated temperature accelerates wear~\cite{Cai2015DataRetention}. A write-heavy
tier therefore degrades far faster than DRAM's effectively unbounded write
endurance.

Third, \emph{power and thermals}: per-stack HBF power is far higher than HBM. H3
assumes roughly 160~W per HBF cube, so six stacks can approach
960~W~\cite{HBFHa2026H3}, potentially exceeding the GPU die's peak power.
Package-local HBF therefore creates a new power and thermal problem rather than
inheriting HBM's modest envelope.

\section{Background}
\label{sec:llm-background}

The preceding section introduced the physical HBF device and its roadmap
organizations. This section explains how inference creates, reuses, and offloads
KV state and why serving systems increasingly move it to a larger tier. It provides
the software and storage context for the HBF deployment evaluated here.

\subsection{KV Cache and Reuse in Agentic Serving}

Agentic and conversational workloads dominate modern LLM deployments:
multi-turn dialogue, tool-use loops, shared system prompts, and
retrieval-augmented generation (RAG). They revisit tokens, reusing prompt prefixes
and session histories across requests. Serving therefore depends less on raw
compute than on memory \emph{capacity} to retain useful prefixes and active
histories and \emph{bandwidth} to restore or stream them without stalling the GPU.

LLM inference has \emph{prefill} and \emph{decode} phases. Prefill processes prompt
tokens in parallel and materializes KV state. A prefix-cache hit avoids recomputing
the matched prompt but still requires locating and delivering its saved blocks to
the GPU\@. Decode generates tokens autoregressively and repeatedly reads the
retained KV history. Each transformer layer stores a key and value vector per
token. For $N$ retained tokens, the unsharded per-request KV footprint is
\begin{equation}
B_{\mathrm{KV/request}}=2NLH_{\mathrm{KV}}D_{\mathrm{head}}b,
\label{eq:kv-request}
\end{equation}
Here, $L$, $H_{\mathrm{KV}}$, $D_{\mathrm{head}}$, and $b$ denote layer count,
KV-head count, head dimension, and bytes per element; the factor of two accounts
for keys and values. Grouped-query and latent-attention designs reduce this
footprint but do not remove the capacity problem.

Unlike static shared weights, KV is generated online, grows during decode, and has
request-, session-, and prefix-specific lifetimes. Stored prefixes matter only if
later requests reuse them before eviction. Continuous batching adds pressure:
admissions consume KV capacity, long requests retain blocks longer, and restoring
offloaded state can delay execution. KV placement, reuse, and request scheduling
must therefore be optimized together.

\subsection{Paged and Pooled KV Storage}

\subsubsection{PagedAttention}
vLLM introduces PagedAttention and block-level KV allocation~\cite{Kwon2023vLLM}.
A block table maps each request's logical KV blocks to non-contiguous physical
blocks in GPU memory. On-demand allocation reduces fragmentation, supports
continuous growth, and enables common-prefix sharing and copy-on-write without a
large contiguous reservation.

Two requests share exact-prefix KV when their token prefixes and model state match.
The benefit grows with matched length but vanishes after eviction. System prompts
and multi-turn dialogue create heavy reuse; independent API calls create little.
The same admission policy can therefore yield very different read/write behavior
across traces.

\subsubsection{Mooncake and distributed KV pools}
Paged allocation solves on-GPU layout but not capacity beyond HBM\@. Mooncake
extends KV into a distributed, KV-cache-centric pool of CPU DRAM, SSDs, and network
resources. It combines a transfer engine with KV-aware, SLO-aware scheduling that
rejects work early under overload~\cite{Kimi2025MoonCake}.
\emph{KV-cache-centric scheduling} treats KV-block location, availability, reuse,
and transfer cost as first-class state beyond request selection.

SSD offloading is asynchronous and outside the write path. New KV lands in
DRAM\@; after DRAM crosses a watermark, a background policy flushes or evicts it to
SSD\@. A master node tracks per-object locations and promotes hot blocks
back~\cite{MooncakeSSDBlog2026}. Flash writes are therefore scheduling decisions,
separate from GPU compute. Save and load move KV in opposite directions:
\begin{align}
\text{save: }&\quad \text{GPU HBM}\rightarrow\text{cache tier},\\
\text{load: }&\quad \text{cache tier}\rightarrow\text{GPU HBM},
\end{align}
Their imbalance is central to the write/read behavior quantified in
Section~\ref{sec:objects}.

\subsection{SSD Offloading of KV}

\subsubsection{Why SSD offloading helps}
HBM and host DRAM are fast but capacity-limited. SSDs add larger, cheaper,
replaceable capacity and retain prefixes over the longer reuse windows needed by
long-context, multi-turn, RAG, and agent workflows.

DeepSeek's API documents a production disk-based context cache with persisted
prefix units and cache-hit/miss token accounting~\cite{DeepSeekDiskCache}.
Production traces show bimodal reuse: most occurs within minutes, but a long tail
returns only after tens of minutes. Recency-only eviction therefore discards blocks
that would still be reused~\cite{MooncakeSSDBlog2026}. SSD offloading is now standard
and provides KV objects for larger tiers to store and restore.

\subsubsection{Modern offload paths}
\begin{figure}[t]
  \centering
  \includegraphics[width=\columnwidth]{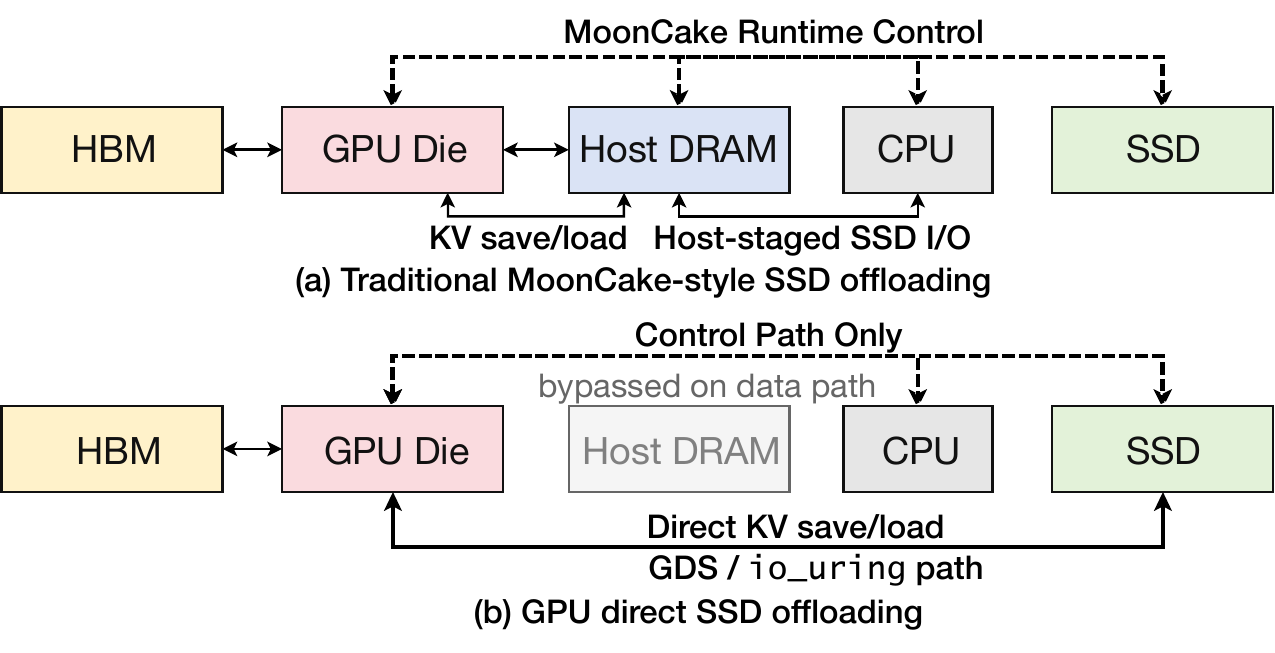}
  \caption{SSD offloading of KV: host-staged path (a) versus GPU-direct path (b).}
  \label{fig:kvcache-ssd}
\end{figure}

In a naive SSD path, data moves through host memory
(GPU~HBM~$\rightarrow$~CPU~I/O~$\rightarrow$~NVMe~$\rightarrow$~SSD).
Submission overhead, extra copies, and fragmented small requests can then dominate
media service (Fig.~\ref{fig:kvcache-ssd}(a)).

Modern paths remove much of this cost (Fig.~\ref{fig:kvcache-ssd}(b)). Linux
\texttt{io\_uring} uses shared queues to reduce system calls and context
switches~\cite{LinuxIoUring}. GPU Direct Storage (GDS) enables direct DMA between
NVMe and GPU memory. Tutti moves object identification and completion to a GPU-side
\texttt{io\_uring} path, removing the CPU from each operation's critical
path~\cite{Tutti2026}.

SSD offloading is therefore a strong, actively optimized baseline. Drives remain replaceable outside the accelerator package, and
their controllers provide buffering, mapping, garbage collection, wear leveling,
and bad-block management.

SSD remains off-package and bounded by the PCIe/NVMe link. HBF provides
a higher-bandwidth, package-local alternative, as described in
Section~\ref{sec:hbf-background}. We evaluate HBF by replacing or
augmenting the SSD backing tier while preserving the KV-object path above.

\section{Evaluation Methodology}
\label{sec:methodology}

\subsection{Simulator and Evaluation Setup}
\begin{table}[t]

  \caption{Evaluation workloads, models, and systems.}
  \label{tab:evaluation}
  \label{tab:traces}
  \label{tab:models}
  \label{tab:modeled-systems}
  \centering
  \scriptsize
  \textbf{(a) Production traces}\\[1pt]
  \setlength{\tabcolsep}{2.2pt}
  \resizebox{\columnwidth}{!}{
  \begin{tabular}{lrrrr}
    \toprule
    Trace & Requests/QPS & Prompt & Output & Multi-turn/reused \\
    \midrule
    \texttt{traceA} & 43,058/5.98 & 1,046/6,436 & 376/808 & 46.3\%/46.8\% \\
    \texttt{traceB} & 172,800/24.0 & 574/1,720 & 39/180 & 0\%/6.0\% \\
    \texttt{coder} & 43,011/5.97 & 4,540/12,776 & 469/1,789 & 38.6\%/39.0\% \\
    \texttt{thinking} & 10,812/1.50 & 3,680/13,115 & 1,666/9,157 & 11.1\%/14.3\% \\
    \bottomrule
  \end{tabular}
  }

  \vspace{3pt}
  \textbf{(b) Model profiles}\\[1pt]
  \resizebox{\columnwidth}{!}{
  \begin{tabular}{llrrrr}
    \toprule
    Model & Arch. & Params & Layers/hidden & Attn./KV heads & Max context \\
    \midrule
    Qwen3-4B & Dense & 4B/4B & 36/2,560 & 32/8 & 40,960 \\
    Qwen3-32B & Dense & 32B/32B & 64/5,120 & 64/8 & 40,960 \\
    DeepSeek-V3.2-685B & MoE & 685B/37B & 61/7,168 & 128/128 & 163,840 \\
    GLM-5.2-753B & MoE & 753B/40B & 78/6,144 & 64/64 & 1,048,576 \\
    Kimi-K2.7-Code-1.1T & MoE & 1.1T/32B & 61/7,168 & 64/64 & 262,144 \\
    \bottomrule
  \end{tabular}
  }

  \vspace{3pt}
  \textbf{(c) Eight-GPU system profiles}\\[1pt]
  \setlength{\tabcolsep}{2.5pt}
  \begin{tabular}{lp{0.32\columnwidth}p{0.38\columnwidth}}
    \toprule
    Profile & Local tier per GPU & Shared secondary tier \\
    \midrule
    SSD/H100 & 96 GB HBM, 3.0 TB/s & 12/24/48 TB SSD; 54.7--218.8 GB/s read \\
    SSD/B200 & 192 GB HBM, 8.0 TB/s & 12/24/48 TB SSD; 54.7--218.8 GB/s read \\
    HBF-1/H100 & 48 GB GDDR7, 1.344 TB/s & 24 TB HBF, 1.2 TB/s \\
    HBF-1/B200 & 96 GB GDDR7, 2.688 TB/s & 48 TB HBF, 2.4 TB/s \\
    HBF-2/H100 & 48 GB HBM3e, 1.5 TB/s & 12 TB HBF, 0.6 TB/s \\
    HBF-2/B200 & 96 GB HBM3e, 4.0 TB/s & 24 TB HBF, 1.2 TB/s \\
    \bottomrule
  \end{tabular}
\end{table}
We use an extended version of TokenSim\footnote{https://github.com/pku-lemonade/TokenSim}~\cite{Wu2025TokenSim}. It models request arrival, prefill, decode, continuous batching, paged KV allocation, admission, preemption, recomputation, KV save/restore, tier transfers, queueing, and request-level latency. Our extensions add Mooncake-style lookup, hashed 16-token KV identifiers, object save/load events, SSD and HBF tiers, per-tier read/write counters, HBF-1/HBF-2 profiles, connector wait accounting, and SLO-oriented outputs. TokenSim's original studies validate its serving behavior against real-machine configurations~\cite{Wu2025TokenSim}; for our extensions, regression checks cover block/byte conservation, save/load ordering, deterministic replay, and per-tier counters. HBF service parameters are sourced or swept. The complete-architecture campaign instantiates the HBF roadmap's HBF-1 (2028) and HBF-2 (2030), with the capacity and bandwidth settings in Table~\ref{tab:evaluation}(c).

We separate media sensitivity from complete architecture comparisons: the media-only view holds GPU organization and policy fixed while varying HBF latency, while complete HBF-1 and HBF-2 change the near tier to the roadmap's GDDR/HBF and HBM/HBF organizations. HBF latency is swept over $8/80$, $12/120$, $20/200$, and $30/300~\mu$s read/write points. The connector uses 16-token hash blocks: prefix lookup matches consecutive identifiers from the prompt start, missing full blocks are saved during prefill after existing-block filtering, and a later matching request restores them before its missing suffix executes.

\subsection{Workloads, Models, and Platforms}

We use the four anonymized production-derived traces listed in Table~\ref{tab:evaluation} from the Aliyun Qwen-Bailian deployment~\cite{Wang2025KVCacheWild}. Each contains a complete two-hour window with fixed request order, lengths, session structure, and 16-token KV identifiers. \texttt{traceA} mixes production traffic with conversational reuse. \texttt{traceB} contains short-output API/text traffic without multi-turn sessions, although shared prefixes still create cross-request reuse. \texttt{coder} contains code-oriented multi-turn context, and \texttt{thinking} combines long reasoning outputs with a smaller multi-turn population. A reused block is a unique hash identifier that appears in more than one request.

The five dense and MoE models listed in Table~\ref{tab:evaluation}(b) supply the architectural parameters used by the current TokenSim profiles. For MoE models, the parameter count reports both total parameters and parameters activated per token.

We model the eight-GPU systems listed in Table~\ref{tab:evaluation}(c) with tensor parallelism 8, pipeline and data parallelism 1, and expert parallelism for MoE models. The SSD tier uses 3.2-TB KIOXIA CM7-V drives with 14.0/6.75~GB/s sequential read/write bandwidth and 3~TB modeled usable capacity per drive~\cite{Kioxia2024CM7V}. The main SSD baseline is the conventional Mooncake connector in the simulator.

We evaluate offered loads from 0.05 to 32 QPS by rescaling request-arrival timestamps while preserving request order, contents, sessions, and KV-reuse relationships.

\subsection{Metrics and Statistical Treatment}

We report TTFT, TBT, end-to-end latency, delivered request and token throughput,
queueing and connector delay, KV hits, blocks and bytes read and written, the
write/read ratio, prefix reuse, recomputation, and SLO goodput. For the
architecture comparison we average each metric over the trace and model campaign
at every system and latency point, because the question is how the complete SSD,
HBF-1, and HBF-2 organizations behave, not how one favorable point behaves. The
write/read ratio and prefix hit rate in Finding~3 come from the per-tier byte
counters over a full two-hour replay of each trace, so they reflect fixed request
contents and ordering rather than a sampled window. Zero-read runs are excluded
from the ratio because it is undefined, and malformed result files are excluded.
Throughout, we label every value as a measured simulator output, a sourced device
parameter, or a modeled projection.

\subsection{Thermal and Endurance Models}

Thermal and endurance models use the serving-trace measurements.
For thermal behavior we build a 16-Hi HBF stack similar to
HBM4, with 128-layer 3D-NAND TLC dies, and solve its steady-state heat flow with
3D-ICE~\cite{Sridhar20103DICE}. We use read and write energy per 16-token KV block,
matching the connector's access granularity, and sweep sustained
single-stack bandwidth to obtain dynamic power and peak temperature. We choose
$80^\circ$C as the model's safe junction temperature and use a throttling
policy that closes the hottest, highest-traffic planes and redirects writes to
cooler planes once the limit is reached. The OCP specification defines normal
operation, light and severe throttling, and shutdown. Its throttling thresholds
are configurable~\cite[\S9]{OCP2026HBFBaseDie}.

For endurance, we normalize the media-write volume from each two-hour,
eight-H100 HBF-2 replay to a daily rate. The capacity-matched SSD reference has
four KIOXIA CM7-V 3.2\,TB drives~\cite{Kioxia2024CM7V}; its vendor 3-DWPD rating
gives $38.4$~TB/day. The HBF tier uses a deliberately favorable TLC endurance
budget with unit write amplification, yielding $21.7$~TB/day for a five-year
target. We compare lifetimes by linear exhaustion of these write budgets under
the measured trace traffic. OCP specifies a product-specific maximum P/E count
and exposes the average P/E count to the host~\cite[\S9]{OCP2026HBFBaseDie}.

\section{A Model of When Faster Flash Helps}
\label{sec:framework}

At matched capacity, an HBF read is faster than an SSD read; that much is
uncontested. The question is whether per-access speedup propagates to end-to-end
request latency, or is absorbed by the rest of the serving pipeline. This is a
system-level question; we formalize it with a single cost--benefit relation.
We express the net change an HBF tier makes to end-to-end latency, relative to the
SSD tier it replaces, as
\begin{multline}
\Delta_{\mathrm{HBF}}=\underbrace{f\,\Delta t_{\mathrm{media}}}_{\text{critical-path I/O}}
-\underbrace{C_{\mathrm{move}}}_{\text{movement/control}}
-\underbrace{C_{\mathrm{pkg}}}_{\text{near-tier opportunity}}\\
-\underbrace{C_{\mathrm{var}}}_{\text{throttle/variability}}
-\underbrace{C_{\mathrm{life}}}_{\text{endurance/capacity}}.
\label{eq:benefit}
\end{multline}
Faster flash improves only the first term, and only in proportion to $f$, the fraction
of end-to-end time spent in \emph{exposed} secondary-tier I/O.
Everything else the HBF tier changes (the near tier it displaces, the traffic it must
absorb, the heat it generates, the wear it takes) enters as a subtracted cost.
The HBF tier is net-positive only when $\Delta_{\mathrm{HBF}}>0$ under target latency,
throughput, power, capacity, and lifetime constraints.
We make this concrete by converting each term into a measurable quantity, from which
three necessary conditions emerge.
Each condition is a \emph{falsifiable prediction}: it states the value a term must
take for HBF to help, while a paired finding measures the value it actually takes
(Table~\ref{tab:framework}).

\noindent\textbf{C1: Read I/O is the serving bottleneck ($f$ must be large).} We decompose
end-to-end time as $T=T_{\mathrm{rest}}+t_{\mathrm{io}}$, where $t_{\mathrm{io}}$ is
the exposed, non-overlapped secondary-tier service and $T_{\mathrm{rest}}$ covers
compute, near-tier service, batching, and queueing; $f=t_{\mathrm{io}}/T$ is the
exposed fraction. To first order, the exposed term scales with per-access latency, so
making the medium $k\times$ faster changes end-to-end time by
\begin{equation}
\frac{\Delta T}{T}=f\Bigl(1-\tfrac{1}{k}\Bigr)\le f .
\label{eq:c1-amdahl}
\end{equation}
The exposed fraction is a hard ceiling on \emph{any} media improvement; even an
infinitely fast tier ($k\!\to\!\infty$) cannot beat $f$.
Equation~\eqref{eq:c1-amdahl} is also invertible, so a media-latency sweep recovers $f$
from a serving measurement directly.
\S\ref{sec:capacity} runs that $k=3.75\times$ sweep, and \S\ref{sec:nmp} asks whether
moving compute into the device can enlarge $f$.

\noindent\textbf{C2: Reads outweigh writes ($\rho>\rho^\star$).} A NAND program is spent the
moment an object enters HBF; that cost is repaid only by later reads that benefit from
the faster medium. Let $\rho$ be the useful read bytes returned per byte written at the
flash tier, $S$ the exposed time each read byte saves, $A$ the write amplification, and
$C$ a per-byte capacity and write-budget cost. Writing is net-positive only when
$\rho S>AC$, i.e.
\begin{equation}
\rho>\rho^\star=\frac{AC}{S}.
\label{eq:object-breakeven}
\end{equation}
Two facts push $\rho^\star$ above one: the write cost is paid unconditionally, while
the repaying reads are speculative, and a single NAND program costs several reads in
energy and wear.
HBF therefore pays off only as a \emph{read-heavy} tier.
For transient KV the hierarchy is expected to deliver the opposite: the near tier retains
the high-$\rho$ hot blocks and hands HBF the write-once tail.
\S\ref{sec:objects} measures the delivered $\rho$, and \S\ref{sec:batching} asks whether
the SSD write optimization that would lower $C$ transfers to HBF.

\noindent\textbf{C3: Sustained bandwidth.} The bandwidth a workload actually receives
is the minimum of interface, array, controller, thermal, and endurance limits
(Eq.~\eqref{eq:effective-bw}), which can fall well below the datasheet peak. Of these
limits, a write-heavy KV stream is most likely to hit the thermal and endurance bounds.
\S\ref{sec:thermal} measures the bandwidth a stack sustains before it overheats, and
\S\ref{sec:cost} measures the write volume it takes over its life.

These conditions are device-agnostic. $f$, $\rho$, and sustained bandwidth
characterize any package-local far tier (CXL-attached flash, other stacked NAND, or
near-data flash), not only HBF.
The rest of the paper tests them on the most demanding object for HBF, \emph{transient
KV} behind an SSD-style save/load connector, pairing two findings with each condition
(Table~\ref{tab:framework}).

\begin{table}[t]
  \caption{Each condition is a falsifiable prediction of the model in
  Eq.~\eqref{eq:benefit}, tested by two findings for transient KV behind an SSD-style
  connector in \S\S\ref{sec:c1}--\ref{sec:c3}.}
  \label{tab:framework}
  \centering

  \setlength{\tabcolsep}{3pt}
  \begin{tabular}{p{0.12\columnwidth}p{0.34\columnwidth}p{0.42\columnwidth}}
    \toprule
    Cond. & Prediction: HBF helps only if & How we test it \\
    \midrule
    C1 & read I/O is the serving bottleneck & media-latency sweep recovers $f$ (F1); base-die NMP tests coverage (F2) \\
    \addlinespace
    C2 & reads outweigh writes ($\rho>\rho^\star$) & per-tier byte counters give delivered $\rho$ (F3); paired batching tests the write cost $C$ (F4) \\
    \addlinespace
    C3 & sustained BW near nominal peak & 3D-ICE thermal sweep (F5); write-endurance budget (F6) \\
    \bottomrule
  \end{tabular}
\end{table}

\section{C1: Read I/O Is the Serving Bottleneck}
\label{sec:c1}

Condition~C1 requires the exposed secondary-tier fraction $f$ to be large. Both findings in this section measure $f$ instead of assuming a value for it.
Finding~1 recovers $f$ from a media sweep over a capacity-matched swap, and Finding~2
asks whether moving the compute into the device can raise it.

\subsection{Finding 1: HBF Trades GPU Capacity and Bandwidth}
\label{sec:capacity}

An HBF read is roughly an order of magnitude faster than an SSD read at matched capacity,
so the swap should make requests faster.
Condition~C1 says that expectation holds only in proportion to $f$,
which no serving study has measured.
Equation~\eqref{eq:c1-amdahl} is invertible, so a media-latency sweep recovers it while a
capacity-matched swap answers what HBF delivers end to end.

Each HBF organization is paired with the SSD baseline of equal secondary-tier capacity,
HBF-1 against SSD24 and HBF-2 against SSD12 (Table~\ref{tab:evaluation}(c)).
Both HBF generations give half of the GPU's near tier to flash and hold equal near-tier
capacity, differing in its technology (GDDR7 vs HBM3e); HBF-1 also carries twice the
flash capacity and twice the flash bandwidth of HBF-2.
The second control varies the medium, moving the read/write point from $30/300$ to
$8/80~\mu$s, the $k=3.75\times$ change Eq.~\eqref{eq:c1-amdahl} needs to recover $f$.
Table~\ref{tab:finding1-capacity} reports both sweeps for every GPU and retained model.

\begin{figure*}[t]
  \centering
  \includegraphics[width=\textwidth]{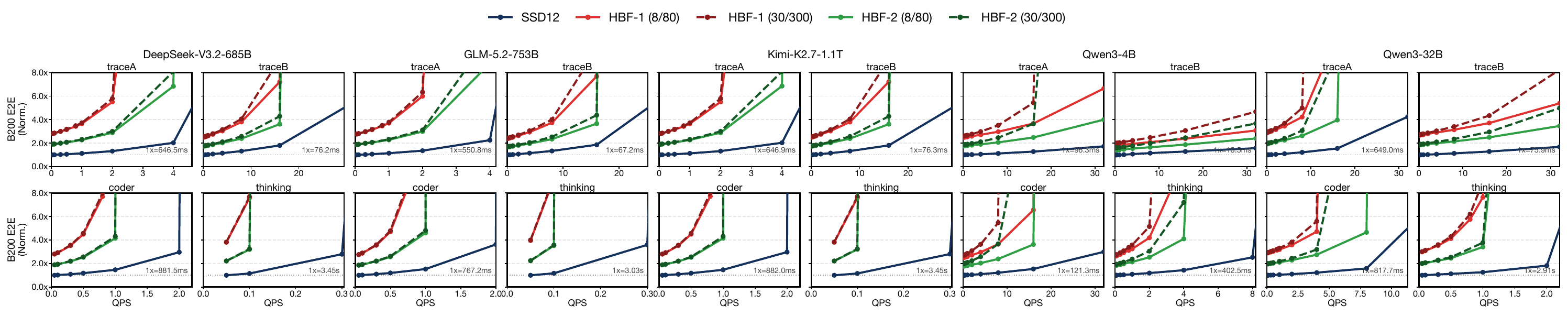}
  \caption{End-to-end latency versus load on B200, normalized within each panel to the
  low-load SSD12 value ($1\times$), across five models and four traces.}
  \label{fig:hbf-e2e}
\end{figure*}

\begin{figure*}[t]
  \centering
  \includegraphics[width=\textwidth]{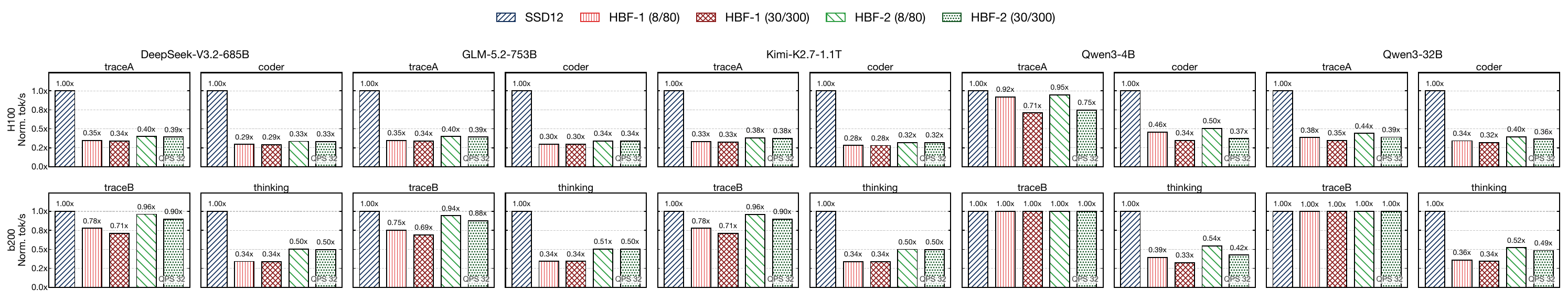}
  \caption{Delivered token throughput normalized to SSD12 ($1\times$) for H100 (top)
  and B200 (bottom), across five models and four traces. }
  \label{fig:hbf-throughput}
\end{figure*}

\begin{figure*}[t]
  \centering
  \includegraphics[width=\textwidth]{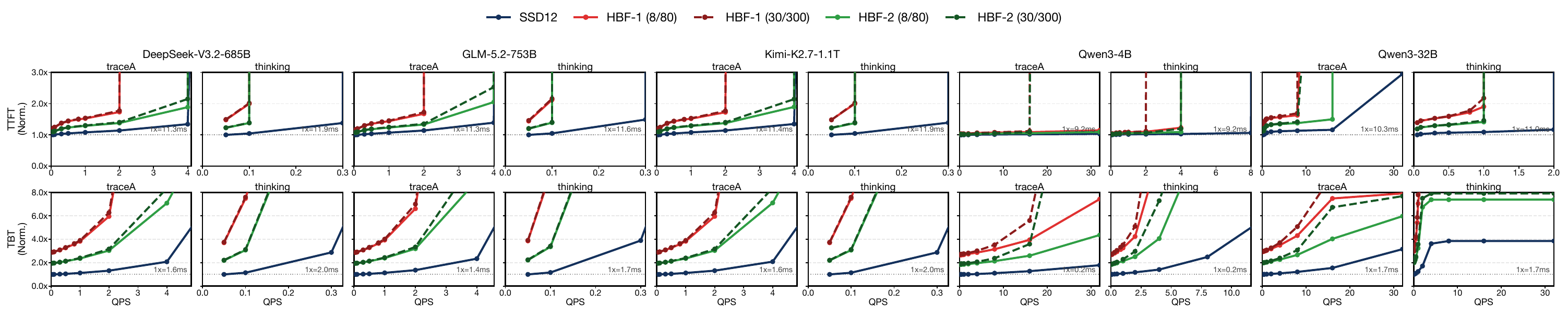}
  \caption{Time-to-first-token (top) and time-between-tokens (bottom) versus load,
  normalized within each panel to the low-load SSD12 value ($1\times$), for
  representative traces on five models.}
  \label{fig:hbf-ttft-tbt}
\end{figure*}

The sweep shows every HBF configuration worse than its SSD pair on every request
metric, with mean end-to-end latency $2$--$5.5\times$ the baseline, delivered
throughput down $4$--$34\%$, and maximum SLO goodput down $1.1$--$2.7\times$; the H100 pairs sit at the
severe end of each range and the B200 pairs at the mild end. Across models and traces every HBF curve rises above
SSD12 and saturates at lower offered load
(Figs.~\ref{fig:hbf-e2e}--\ref{fig:hbf-ttft-tbt}).

The media sweep moves almost nothing. A $k=3.75\times$ improvement in raw media latency
shifts mean end-to-end latency by $0.75\%$ (HBF-1) and $0.88\%$ (HBF-2), which inverts
through Eq.~\eqref{eq:c1-amdahl} to
$f=(\frac{\Delta T}{T})/(1-\frac{1}{k})\approx1\%$. The media change remains measurable
elsewhere, moving throughput ${\sim}5\%$ and SLO goodput $14$--$23\%$.

\begin{figure}[t]
  \centering
  \includegraphics[width=\columnwidth]{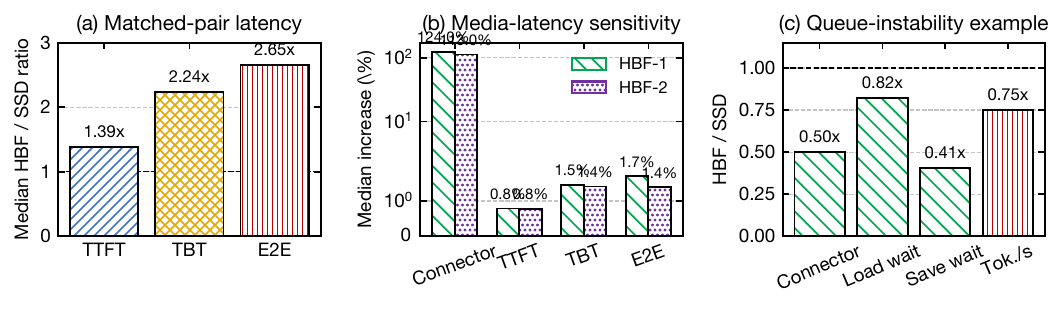}
  \caption{A faster connector, a slower system: HBF cuts transfer time yet
  raises end-to-end latency because the smaller near tier destabilizes the queue.}
  \label{fig:headline-results}
\end{figure}

\noindent\textbf{Why the faster medium loses.}
Media service does not reach the request's critical path.
On GLM-5.2, \texttt{traceA}, B200 at $32$~QPS, HBF-1's cumulative connector service time
across all requests is $3.2\times$ \emph{lower} than SSD's ($399$ vs $1{,}293$~s), yet its
average per-request end-to-end latency over the replay is $2.6\times$ \emph{higher}
($4{,}386$ vs $1{,}682$~s) and its TBT doubles (Fig.~\ref{fig:headline-results}).
Across the campaign $4{,}741$ matched pairs share that shape,
so the bottleneck lies outside media service.

Two comparisons locate it, and neither is a single-variable control.
The HBF/SSD pair changes near-tier capacity and bandwidth together, halving both ($96$ to
$48$~GB and $3.0$ to $1.5$~TB/s on H100), so it prices the package trade without
separating its two parts. The HBF-1/HBF-2 pair instead holds near-tier capacity fixed, and
HBF-1 is $16$--$64\%$ slower end to end despite carrying twice the flash capacity and
bandwidth, so a strictly better flash tier does not recover the loss. What the swap
degrades is the near tier that governs how many requests the server admits at once.

The result is a ceiling, not a tuning outcome. With $f\approx1\%$, a perfect
zero-latency medium could remove at most one percent of the critical path, while the
package trade that buys it costs $2$--$5.5\times$ in end-to-end latency. In
Eq.~\eqref{eq:benefit}, $C_{\mathrm{pkg}}$ exceeds $f\,\Delta t_{\mathrm{media}}$ by two
orders of magnitude. One alternative explanation is that $f$ is small only because the
compute sits far from the data, which Finding~2 tests directly.

\begin{table*}[t]
  \caption{B200 HBF performance relative to capacity-matched SSD baselines.
  HBF-1 is normalized to SSD24 and HBF-2 to SSD12. Positive latency changes
  indicate slowdowns; negative throughput and goodput changes indicate losses.}
  \label{tab:finding1-capacity}
  \centering
  \begingroup
  \normalsize

  \begin{tabular}{l rr rr rr rr}
    \toprule
    Model & \multicolumn{2}{c}{$\Delta$TTFT}
      & \multicolumn{2}{c}{$\Delta$E2E Latency}
      & \multicolumn{2}{c}{$\Delta$Throughput}
      & \multicolumn{2}{c}{$\Delta$Goodput} \\
    \cmidrule(lr){2-3}\cmidrule(lr){4-5}\cmidrule(lr){6-7}\cmidrule(lr){8-9}
    & HBF-1 & HBF-2 & HBF-1 & HBF-2 & HBF-1 & HBF-2 & HBF-1 & HBF-2 \\
    \midrule
    Kimi-K2.7
      & $+96.6\%$ & $+44.7\%$ & $+285.2\%$ & $+131.4\%$ & $-24.3\%$ & $-12.7\%$ & $-51.7\%$ & $-41.7\%$ \\
    GLM-5.2
      & $+94.3\%$ & $+45.9\%$ & $+280.0\%$ & $+133.2\%$ & $-24.6\%$ & $-13.7\%$ & $-37.5\%$ & $-12.5\%$ \\
    DeepSeek-V3.2
      & $+96.5\%$ & $+44.7\%$ & $+284.7\%$ & $+131.3\%$ & $-24.2\%$ & $-12.7\%$ & $-51.7\%$ & $-41.7\%$ \\
    Qwen3-32B
      & $+73.4\%$ & $+42.1\%$ & $+267.4\%$ & $+136.0\%$ & $-11.8\%$ & $-7.9\%$ & $-40.0\%$ & $-25.1\%$ \\
    Qwen3-4B
      & $+19.2\%$ & $+14.6\%$ & $+184.1\%$ & $+103.4\%$ & $-5.2\%$ & $-3.6\%$ & $-46.3\%$ & $-35.9\%$ \\
    \bottomrule
  \end{tabular}
  \endgroup
\end{table*}

\begin{findingtakeaway}{Finding 1}
Replacing SSD with HBF sacrifices the near-tier capacity and bandwidth that govern admission, so faster flash does not improve end-to-end serving.
\end{findingtakeaway}

\subsection{Finding 2: Base-Die Compute Cannot Raise the Exposed Fraction}
\label{sec:nmp}

Finding~1 leaves $f$ near $1\%$.
One reading of that number is that the compute sits too far from the data,
in which case moving the compute to the data should raise $f$.
The standard form of that move is a near-memory attention engine on the HBF base die,
as HBM-PIM does for DRAM.
We ask whether it does.

We model an additional decode-attention engine on the base die under assumptions chosen to favor HBF.
It is charged no area, power, or programmability penalty,
and every eligible token hits at the device service time,
so a negative result is a lower bound.
The two modes differ in reach.
The HBM engine accelerates all HBM-resident KV;
the HBF engine reaches only KV already resident in HBF,
and hot HBM-resident KV falls back to GPU attention.
Table~\ref{tab:nmp-setup} lists the full configuration.

\begin{table}[t]
  \caption{Finding~2 NMP configuration.}
  \label{tab:nmp-setup}
  \centering

  \setlength{\tabcolsep}{5pt}
  \begin{tabular}{ll}
    \toprule
    Parameter & Value \\
    \midrule
    Model / cluster & DeepSeek-V3.2-685B, $8\times$H100 (TP8, DP1) \\
    Memory preset & HBF2, $12$~TB shared HBF tier \\
    HBF latency & $12~\mu$s read, $120~\mu$s write \\
    Modes & HBF base-die NMP, HBM base-die NMP \\
    NMP type & 24 FP16 MAC arrays at 600~MHz \\
    Internal-stack \\ NMP bandwidth & 896~GB/s \\
    QPS tiers & 0.05, 0.1, 0.2, 0.5, 1, 2, 4, 8, 16, 32 \\
    \bottomrule
  \end{tabular}
\end{table}

\begin{figure}[t]
  \centering
  \includegraphics[width=\columnwidth]{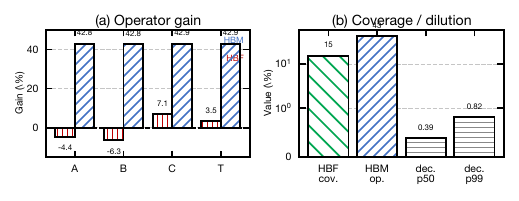}
  \caption{Base-die NMP operator leverage, HBF coverage, and end-to-end dilution. A--T
  denote traceA, traceB, coder, and thinking.}
  \label{fig:nmp}
\end{figure}

Fig.~\ref{fig:nmp}(a) reports both engines.
Averaged over the $40$ sweep points, the HBM engine cuts decode-attention time by
$42.84\%$, and it is stable across the whole sweep ($42.78$--$42.86\%$).
The HBF engine averages $-0.03\%$ on the same basis, that is, no benefit at all.
Individual points swing from $-29.05\%$ to $+20.49\%$,
and it is \emph{slower} than GPU attention at $16$ of $40$ points,
turning positive only near saturation.
The end-to-end effect is smaller again.
Replacing the HBF engine with the much stronger HBM engine moves decode p50 by $0.39\%$
and decode p99 by $0.82\%$ on non-saturated points.

Two quantities account for this behavior.
The first is \emph{coverage}.
A base-die engine sees only the cold KV the runtime has already placed in HBF,
and the mean HBF-resident fraction is just $15.48\%$ (Fig.~\ref{fig:nmp}(b)),
so even a perfect HBF engine cannot touch the common HBM-resident case.
Table~\ref{tab:nmp-pertrace} tracks residency and leverage across traces.
\texttt{thinking} keeps only $4\%$ of attention in HBF and sees essentially no benefit.
\texttt{traceA} at $27\%$ residency has the most eligible work,
yet it is still net-negative, which brings in the second quantity, \emph{scale}.
The HBF path pays a $12~\mu$s access on every step,
and many low-load decode steps are too small to amortize it.
The end-to-end numbers follow the same accounting,
since MoE compute, expert communication, prefill, HBF transfers, and queueing are
unchanged by either engine.

\begin{table}[t]
  \caption{Per-trace NMP leverage and coverage. }
  \label{tab:nmp-pertrace}
  \centering

  \setlength{\tabcolsep}{4pt}
  \begin{tabular}{lrrrr}
    \toprule
    Trace & HBF-NMP & HBF-NMP & HBM-NMP & HBF \\
     & mean & QPS$\le$1 & mean & residency \\
    \midrule
    \texttt{traceA}   & $-4.43\%$ & $-21.67\%$ & $42.84\%$ & $26.62\%$ \\
    \texttt{traceB}   & $-6.27\%$ & $-14.11\%$ & $42.81\%$ & $13.63\%$ \\
    \texttt{coder}    & $7.13\%$  & $-1.68\%$  & $42.85\%$ & $17.64\%$ \\
    \texttt{thinking} & $3.45\%$  & $1.96\%$   & $42.85\%$ & $4.02\%$  \\
    \bottomrule
  \end{tabular}
\end{table}

Moving compute into the device raises neither term of $f$.
It enlarges neither the resident fraction nor the per-step work, so C1 survives the
intervention.
The conclusion is not that base-die compute is useless but that it is the wrong lever
for this bottleneck.
It accelerates a small, cold, microsecond-latency slice, so base-die silicon is better
spent on the control plane than on a general attention engine.
The next condition turns to what a swap costs on the write side.

\begin{findingtakeaway}{Finding 2}
Base-die attention compute cannot overcome HBF's limited KV coverage, so its silicon is better spent on control rather than general attention.
\end{findingtakeaway}

\section{C2: Reads Outweigh Writes}
\label{sec:c2}

Condition~C2 requires HBF to be a read-heavy tier,
$\rho>\rho^\star\!\gtrsim\!1$ (Eq.~\eqref{eq:object-breakeven}).
Finding~3 measures the reads-per-write the hierarchy delivers,
and Finding~4 asks whether the standard SSD write optimization can lower the write cost $C$.

\subsection{Finding 3: SSD-Style Hierarchy Makes HBF Write-Heavy}
\label{sec:objects}

Finding~1 showed the flash service sits off the critical path.
Whether it earns its place is a separate question.
Every byte HBF serves must first be written there,
so C2 asks whether those writes are repaid by later reads.
A hierarchy tuned for reuse could land on either side of that threshold.

We replay each of the four production traces for two hours
and measure the write and read byte counts at the flash tier directly.
Their ratio converts to the delivered $\rho$ that C2 tests.
We also record the prefix hit rate,
which separates cache effectiveness from flash-tier traffic.

Table~\ref{tab:finding3-writeheavy} reports all three quantities.
Every trace writes more to the flash tier than it reads back,
from 1.14$\times$ (\texttt{coder}) to 4.90$\times$ (\texttt{traceB}),
so $\rho$ ranges from 0.20 to 0.88,
below break-even on every trace and several-fold below on the write-heaviest.
The prefix hit rate is 38--53\%,
so reuse is found even where the flash tier reads little back.

\begin{table}[h]
  \caption{Flash-tier traffic over a full two-hour replay.}
  \label{tab:finding3-writeheavy}
  \centering

  \setlength{\tabcolsep}{4pt}
  \begin{tabular}{lrrrr}
    \toprule
    Metric & \texttt{traceA} & \texttt{traceB} & \texttt{coder} & \texttt{thinking} \\
    \midrule
    Write/read ratio & $2.14\times$ & $4.90\times$ & $1.14\times$ & $2.20\times$ \\
    $\rho$ (reads/write) & $0.47$ & $0.20$ & $0.88$ & $0.45$ \\
    Prefix hit rate & 38.21\% & 52.46\% & 45.58\% & 44.12\% \\
    \bottomrule
  \end{tabular}
\end{table}

\noindent\textbf{Reuse distribution across blocks.}
The two observations reconcile through the reuse distribution, not the average.
Reuse is extremely skewed.
The top decile of unique blocks supplies $64\%$ (\texttt{traceA})
to $95$--$100\%$ (\texttt{thinking}, \texttt{traceB}) of all reuse reads,
with Gini $0.80$--$1.0$.
Those hot blocks are the ones the near tier keeps resident,
so those hits are served from HBM or GDDR,
and what reaches HBF is the long cold tail, written once and rarely read again.
\texttt{traceB} is the extreme, with only $6\%$ of blocks ever reused,
so almost everything it sends HBF is write-once
and its ratio is highest.
A cache tier that mostly writes is therefore the hierarchy \emph{doing its job},
not a misconfiguration.
It keeps reuse where it is fastest to serve
and hands the tier that should hold read-mostly data a write-mostly stream.

Under save/restore pooling, $\rho$ depends on both object class and placement.
Stable weights and static tensors support many reads per write ($\rho\gg1$),
and shared-prefix KV can amortize one write over many requests.
Private decode KV grows token by token and dies when the request finishes,
leaving few opportunities for a later restore.
The traces quantify cross-request reuse:
The fraction of blocks ever reused across requests
is $46.8\%$, $6.0\%$, $39.0\%$, and $14.3\%$.
The pooling policy determines how much of that reuse reaches the flash tier.
This is why H3 and HAVEN place stable, read-mostly objects
in HBF~\cite{HBFHa2026H3,HBFHsu2026HAVEN}.
The problem is not that HBF cannot hold KV,
but that the SSD-style hierarchy routes the least reusable KV
to the device least able to absorb writes.

\begin{findingtakeaway}{Finding 3}
The two-tier hierarchy keeps reusable KV in the near tier and sends HBF a write-heavy cold stream whose writes are rarely repaid.
\end{findingtakeaway}

\subsection{Finding 4: SSD Write-Batching helps little for HBF}
\label{sec:batching}

Finding~3 establishes $\rho$ under the unchanged pooling policy,
so the remaining lever in Eq.~\eqref{eq:object-breakeven} is the write cost $C$.
SSD KV serving already has the standard tool for lowering it.
Mooncake-style batching coalesces many small KV writes into large buckets,
amortizing the drive's slow per-operation I/O
and turning scattered writes into near-sequential transfers.
Whether that win is a property of the optimization or of the medium it was built for
decides whether it carries onto HBF.

We run the identical bucketing mechanism on both media
over a shared \texttt{traceA}/\texttt{traceB} matrix,
holding HBF's random and sequential bandwidth \emph{equal}
so batching can help only by amortizing fixed write-operation latency.
We compare three arms, no-batch, the Mooncake bucketed path,
and a single-key control that pays the bucketing cost without grouping anything.

\begin{figure}[t]
  \centering
  \includegraphics[width=\columnwidth]{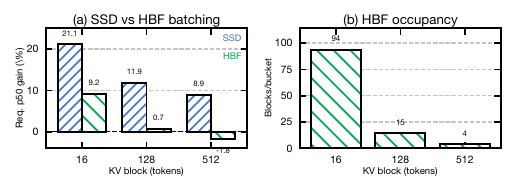}
  \caption{Batching gain does not transfer from SSD to HBF. (a)~Request-p50 reduction
  from Mooncake batching on SSD-12 vs.\ HBF-2 across KV block sizes; SSD gains hold
  everywhere while HBF fades to negative. (b)~HBF-2 bucket occupancy collapses as blocks
  grow, removing the only latency HBF batching can amortize.}
  \label{fig:batching}
\end{figure}

Table~\ref{tab:finding4-batching} and Fig.~\ref{fig:batching}(a) give the paired result.
On SSD-12, request p50 falls by $8.9$--$21.1\%$ at every block size.
The same mechanism on HBF-2 recovers $2.69\%$ overall, $5.2\times$ smaller,
and the benefit is concentrated at the smallest block,
$+9.2\%$ at $16$ tokens, fading to $+0.72\%$ at $128$
and turning \emph{negative} ($-1.85\%$) at $512$.
The SSD gain is flat across block size; the HBF gain decays to zero and then costs.

\noindent\textbf{Composition of the write cost.}
The contrast follows from what each medium's write cost is made of.
On SSD a write is dominated by high fixed per-operation overhead
and a random-access penalty, and usable bandwidth is low relative to demand.
Batching attacks all three at once, so its win is large and survives coarse blocks.
Our HBF service model isolates fixed-overhead amortization:
random and sequential bandwidth are equal, so bucketing gains no
scattered-to-sequential bandwidth advantage.
The only lever left is amortizing fixed operation latency,
which pays off only when a bucket is dense with tiny blocks,
about $94$ blocks per bucket at $16$ tokens but $4$ at $512$
(Fig.~\ref{fig:batching}(b)).
The single-key control makes that residue explicit.
With the batching path enabled but no grouping it is \emph{worse} than no-batch
on every metric, because it pays bucketing overhead for no amortization.

SSD-style buckets amortize fixed I/O overhead. OCP specifies
complete 4\,KiB page writes and sequential programming. For maximum parallelism,
banks within a die use the same page number~\cite[\S\S5.4, 11.2.2]{OCP2026HBFBaseDie}.
Larger buckets preserve the write-heavy stream. Their measured benefit is too
small to rescue C2.

\begin{table}[t]
  \caption{Paired batching benefit (batched vs.\ matched no-batch), SSD-12 vs.\ HBF-2
  on the shared \texttt{traceA}/\texttt{traceB} matrix. }
  \label{tab:finding4-batching}
  \centering
  \setlength{\tabcolsep}{4pt}
  \begin{tabular}{lrrrr}
    \toprule
    & \multicolumn{2}{c}{Req.\ p50 gain} & \multicolumn{2}{c}{Token/s gain} \\
    \cmidrule(lr){2-3}\cmidrule(lr){4-5}
    Block (tokens) & SSD-12 & HBF-2 & SSD-12 & HBF-2 \\
    \midrule
    overall  & $+13.98\%$ & $+2.69\%$ & $+2.04\%$ & $+0.31\%$ \\
    16       & $+21.14\%$ & $+9.20\%$ & $+3.08\%$ & $+0.91\%$ \\
    128      & $+11.88\%$ & $+0.72\%$ & $+1.56\%$ & $+0.06\%$ \\
    512      & $+8.92\%$  & $-1.85\%$ & $+1.48\%$ & $-0.04\%$ \\
    \midrule
    \multicolumn{5}{l}{\textbf{HBF-2 controls (no SSD analogue)}} \\
    \quad batched, all blocks & --- & $+1.76\%$ & --- & $+0.31\%$ \\
    \quad 1-block control     & --- & $-4.25\%$ & --- & $-0.37\%$ \\
    \bottomrule
  \end{tabular}
\end{table}

\begin{findingtakeaway}{Finding 4}
SSD-style bucketing delivers little benefit under the evaluated HBF service model and leaves the write-heavy pooling stream unchanged.
\end{findingtakeaway}

\section{C3: Read/Write Performance Sustains}
\label{sec:c3}

Condition~C3 requires the bandwidth the workload receives to stay near the datasheet
peak.
Finding~5 measures what an HBF stack sustains at run time,
and Finding~6 measures what the write side costs over the device's life.

\subsection{Finding 5: Write-Heavy Traffic Overheats HBF and Throttles Bandwidth}
\label{sec:thermal}

Finding~3 established that the flash stream a KV hierarchy generates is write-heavy, and
writes cost more power than reads.
What sustained bandwidth an HBF stack can hold under that traffic is therefore open;
the interface peak bounds it but does not answer it.

\begin{figure}[t]
  \centering
  \includegraphics[width=\columnwidth]{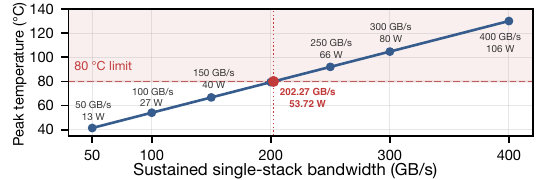}\\[0.2ex]
  \includegraphics[width=\columnwidth]{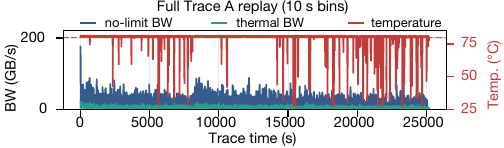}\\[0.2ex]
  \includegraphics[width=\columnwidth]{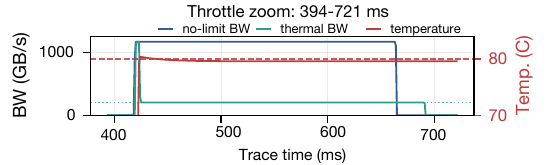}\\[0.2ex]
  \caption{(a) Dynamic power and peak temperature vs.\ single-stack bandwidth. (b) Full replay: thermal control caps delivered bandwidth. (c) Throttle zoom: temperature stays near $80^\circ$C while bandwidth is capped.}
  \label{fig:thermal}
\end{figure}

\noindent\textbf{Thermal model and throttling policy.}
A 3D-ICE~\cite{Sridhar20103DICE} model of a 16-Hi TLC stack, driven by the access energy
of a 16-token KV block, sweeps sustained single-stack bandwidth (\S\ref{sec:methodology}).
The thermal policy in Section~\ref{sec:methodology} closes the hottest,
highest-traffic planes and redirects writes to cooler ones.

Dynamic power scales linearly with bandwidth, and the stack hits the $80^\circ$C limit
at only $202.27$~GB/s and $53.72$~W, far below interface peak, so it cannot run
continuously above that point (Fig.~\ref{fig:thermal}(a)).
In the replay, the unconstrained run holds peak bandwidth through a burst, while the
thermal controller drops the constrained run to the $202$~GB/s safe point and holds
temperature near $80^\circ$C (Fig.~\ref{fig:thermal}(b)--(c)).

\noindent\textbf{Why the stack runs hot.}
Delivered bandwidth is not a constant.
It varies in \emph{time} as hot regions throttle and cool, and in \emph{space} as the
runtime steers traffic away from the hottest planes.
The cause traces back to Finding~3.
Because reuse stays in the near tier and writes dominate the flash stream at higher power
per access, the device is pushed to its limit by exactly the write-heavy traffic the
hierarchy sends it.
A read-mostly workload at the same bandwidth would run cooler.

A scheduler that assumes a fixed flash bandwidth will mispredict service time and build
queues it did not expect.
The OCP specification likewise treats thermal state as part of the host contract:
severe throttling halts data I/O while retaining maintenance service~\cite[\S9.2]{OCP2026HBFBaseDie}.
The unchanged pooling runtime therefore needs more than a peak-bandwidth number
to sustain its write-heavy stream.

\begin{findingtakeaway}{Finding 5}
Write-heavy KV traffic drives HBF into thermal throttling, making delivered bandwidth lower and less predictable than its rated peak.
\end{findingtakeaway}

\subsection{Finding 6: HBF Wears Out Sooner Under Write-Heavy KV Pooling}
\label{sec:cost}

Finding~5 measured what the write-heavy stream costs in the moment, when its power
throttles delivered bandwidth.
The same stream also has a cumulative cost.
Every write spends part of a finite program/erase budget. We compare the
trace-measured write volume against the HBF and capacity-matched SSD endurance
budgets defined in Section~\ref{sec:methodology}.

\begin{figure}[t]
  \centering
  \includegraphics[width=\columnwidth]{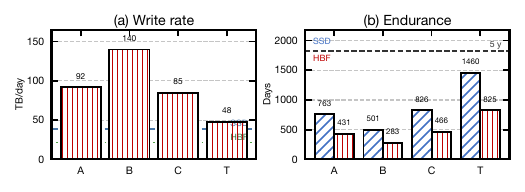}
  \caption{KV write volume and endurance budget. A--T denote traceA, traceB, coder, and thinking.}
  \label{fig:endurance}
\end{figure}

Every trace writes past \emph{both} envelopes ($48$--$140$~TB/day), so under linear TBW
exhaustion the nominal HBF tier lasts a constant $0.56\times$ the SSD pool's life across
all four traces (Fig.~\ref{fig:endurance}).
Coarsening KV blocks from $16$ to $128$ to $512$ tokens raises the apparent prefix hit rate
yet \emph{increases} HBF write volume from $188$ to $237$ to $258$~TB ($1.38\times$).

\noindent\textbf{Why the shortfall is structural.}
HBF is attached like memory but behaves like flash.
OCP assigns ECC, bad-block handling, and write accumulation to the base die.
Wear leveling can run in the base die or on the host~\cite[\S\S4.4, 5.4, 11.4]{OCP2026HBFBaseDie}.
Wear leveling distributes wear but leaves the pooling stream's programming
demand intact. Unlike an SSD's internal reclamation machinery~\cite{FlashFTLSurvey},
OCP delegates live-data movement and reclamation to the host. Zone remapping
requires invalid data and quiesced reads before rewriting~\cite[\S11.4]{OCP2026HBFBaseDie}.
OCP specifies host refresh for data age and read-disturb counts, and
24-hour powered-on retention at $85^\circ$C~\cite[\S\S9, 11.5]{OCP2026HBFBaseDie}.
These responsibilities make write volume, wear distribution, and data lifetime
explicit inputs to placement, rather than properties hidden by a peak-bandwidth number.
Block granularity does not route around that state.
Small blocks maximize the number of write operations. Larger blocks increase
bytes written and write amplification: partial reuse can still require
rewriting the entire block.
Both keep the device on the wrong side of Eq.~\eqref{eq:object-breakeven}.

Our trace-driven endurance comparison shows that HBF wears out sooner than the
capacity-matched SSD pool it replaces, exposing the endurance cost
$C_{\mathrm{life}}$ in Eq.~\eqref{eq:benefit}.
Writing short-lived KV into HBF pays twice: programming must complete, and
programs consume the endurance budget.

\begin{findingtakeaway}{Finding 6}
Under unchanged SSD-style KV pooling, HBF wears out sooner than the capacity-matched SSD pool: the measured write-heavy stream exhausts its endurance budget faster.
\end{findingtakeaway}

\section{Discussion and Related Work}
\label{sec:discussion}

\subsection{Synthesis}

\begin{figure}[t]
  \centering
  \includegraphics[width=\columnwidth]{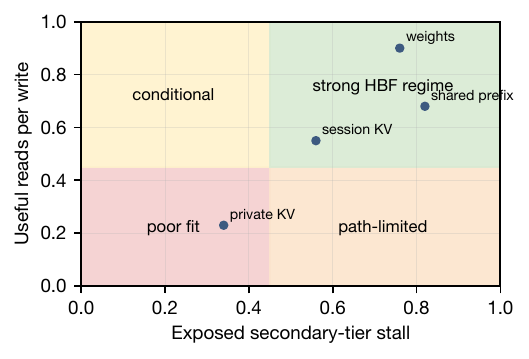}
  \caption{Placement map implied by the findings: exposed stall (C1) against
  reads-per-write (C2), with sustainable bandwidth (C3) shifting the boundary.}
  \label{fig:regime-map}
\end{figure}

The six findings falsify the three conditions of Section~\ref{sec:framework} in turn, and
together they mark where HBF does help.
An object belongs in HBF only when its restore delay is the serving bottleneck (C1),
its reads outweigh its writes ($\rho>\rho^\star$, C2), and the bandwidth it needs is
sustainable under the package's power and endurance budget (C3).
Fig.~\ref{fig:regime-map} places representative object classes on the two axes that decide
this (exposed stall and reads-per-write), with sustainable bandwidth shifting the boundary.
Transient decode KV sits in the low-exposure, low-$\rho$ corner where all three conditions
fail; weights and shared prefixes sit in the opposite corner, which is where H3 and HAVEN
place their HBF-resident state.

The three conditions use device-agnostic quantities: exposed critical-path fraction,
reads per write, and sustained bandwidth.
The same map therefore applies to other package-local far tiers such as CXL-attached or
near-data flash, not only to HBF.
The map gives the axes and the break-even rule of Eq.~\eqref{eq:object-breakeven}
rather than a calibrated boundary across the latency$\times$bandwidth plane.
A working runtime requires critical-path-aware access, reuse-aware admission,
and thermal-aware write budgeting. The OCP specification reinforces this
object-aware direction by describing separate weight and KV channels to manage
endurance and capacity utilization~\cite[\S13.3.3]{OCP2026HBFBaseDie}.
Our measurements establish why an unchanged SSD-style pool is a poor match.

\subsection{Related Work}

\noindent\textbf{KV-cache management.} PagedAttention enables block allocation and prefix
sharing~\cite{Kwon2023vLLM}, quantization and compaction shrink KV
size~\cite{Liu2024ICMLKIVI,Hooper2024NeurIPSKVQuant,Zhang2025OSDIDiffKV}, and Prompt
Cache and CachedAttention reuse state across prompts~\cite{Gim2024promptcache,Bin2024CachedAttention}.
Mooncake, LMCache, Strata, and Tutti offload KV across CPU, SSD, and network tiers
with GPU-assisted I/O~\cite{Kimi2025MoonCake,LMCache2025,Strata2025,Tutti2026}, and
DeepSeek's disk context cache evidences persisted prefix reuse~\cite{DeepSeekDiskCache}.
We ask how this SSD-offload stack behaves when its backing medium moves into the
accelerator package.

\noindent\textbf{HBF.} Prior studies explore HBF as either a read-mostly extension
or the main GPU memory. H3 places read-only weights and shared precomputed KV in
HBF while retaining generated state in HBM~\cite{HBFHa2026H3}; HAVEN stores a vector
collection in HBF and reranks near the device~\cite{HBFHsu2026HAVEN}. FlashAccel
adds six HBF stacks to an HBM-based GPU and reports throughput $2.54\times$ that
of its HBM-only baseline~\cite{Wang2026FlashAccel}.
The POSTECH study evaluates both one HBM plus seven HBF stacks and an all-HBF
configuration with eight HBF stacks. Weights and KV reside in flash; intermediate
data use the retained HBM or base-die SRAM, respectively~\cite{Son2026ExploringHBF}.
These configurations explore HBF as primary memory, distinct from the HBF
roadmap's GDDR--HBF and shared HBM--HBF organizations evaluated here. Our focus is
the HBF-1/HBF-2 package tradeoff under an SSD-style KV hierarchy and production-trace
replay. The studies thus address different organizations, data paths, and serving
conditions.

\noindent\textbf{Flash reliability.} SSD controllers manage mapping, reclamation,
and wear~\cite{FlashFTLSurvey}.
The OCP specification divides management between the base die and the host, which handles data
movement and refresh~\cite[\S\S11.4--11.5]{OCP2026HBFBaseDie}. NAND retention
depends on wear and temperature~\cite{Cai2015DataRetention}.

\subsection{Limitations}

The hardware profiles follow the HBF roadmap's near-term organizations and
the service parameters in Section~\ref{sec:methodology}. The architecture
comparison measures the joint effect of the near tier and backing tier under
the same pooling policy; the latency sweep isolates media-service sensitivity.
Thermal and endurance results use the stated temperature policy and TLC write
budget, while cost is expressed through package-resource trade-offs.
We take the host/device contract directly from OCP to explain the layout,
write-management, and sustained-service requirements of this deployment.

\FloatBarrier

\section{Conclusion}
\label{sec:conclusion}

For the HBF roadmap organizations studied here---HBF-1 (2028) and HBF-2 (2030)---
replacing an SSD KV tier with HBF does not improve LLM serving; under the same
Mooncake-style runtime it makes serving worse. Our model explains why: a faster far tier
helps only when read I/O is the serving bottleneck (C1), reads outweigh writes (C2), and
its bandwidth is sustainable (C3), and all three fail for transient
KV behind an SSD-style connector.
HBF sucks as a drop-in SSD KV pool under this unmodified runtime: serving slows
down, and the trace-driven endurance comparison shows that HBF wears out sooner
than the capacity-matched SSD pool it replaces. The OCP specification defines
page-write and layout rules and lifecycle controls. These reinforce the need
to change the pooling policy and data path, not merely replace the backing medium.

\bibliographystyle{ACM-Reference-Format}
\bibliography{reference}

\end{document}